\documentclass[prd,preprintnumbers,floatfix,nofootinbib,notitlepage]{revtex4}

\usepackage{latexsym}
\usepackage{epsfig}
\usepackage{amssymb}
\usepackage{amsmath}

\newcommand{\ba}{\begin{eqnarray}}
\newcommand{\ea}{\end{eqnarray}}
\newcommand{\be}{\begin{equation}}
\newcommand{\ee}{\end{equation}}

\newcommand{\al}{\alpha}
\newcommand{\bt}{\beta}
\newcommand{\ga}{\gamma}

\begin{document}

%\title{Nonlocal gravity from generalised spacetime structure}
\title{Effectively nonlocal metric-affine gravity}

\author{Alexey Golovnev}
\affiliation{Faculty of Physics, St. Petersburg State University, Ulyanovskaya ul., d. 1, Saint Petersburg 198504, Russia }
\email{agolovnev@yandex.ru}

\author{Tomi Koivisto}
\affiliation{Nordita, KTH Royal Institute of Technology and Stockholm University, Roslagstullsbacken 23, SE-10691 Stockholm, Sweden}
\email{tomik@astro.uio.no}

\author{Marit Sandstad}
\affiliation{Nordita, KTH Royal Institute of Technology and Stockholm University, Roslagstullsbacken 23, SE-10691 Stockholm, Sweden}
\affiliation{Institute for Theoretical Astrophysics, University of Oslo, P.O. Box 1029 Blindern, N-0315 Oslo, Norway}
\email{marit.sandstad@astro.uio.no}

\date{\today}

\begin{abstract}

In metric-affine theories of gravity such as the C-theories, the spacetime connection is associated to a metric that is nontrivially related to the physical metric.
In this article, such theories are rewritten in terms of a single metric and it is shown that they can be recast as effectively nonlocal gravity. With some assumptions, known ghost-free theories with non-singular and cosmologically interesting properties may be recovered. Relations between different formulations are analysed at both perturbative and nonperturbative levels taking carefully into account subtleties with boundary conditions in the presence of integral operators in the action, and equivalences between theories related by nonlocal redefinitions of the fields are verified at the level of equations of motion. This suggests a possible geometrical interpretation of nonlocal gravity as an emergent property of non-Riemannian spacetime structure.

\end{abstract}

\preprint{NORDITA-2015-116}

\maketitle

\section{Introduction}

There are two basic ways of modifying gravity: considering either more general action for the metric or allowing spacetime degrees of freedom besides the metric. 

In General Relativity (GR), gravity can be interpreted as the curvature of spacetime, and is described within the framework of Riemannian geometry in terms of the metric, a tensor field that furthermore in GR obeys second order equations of motion. In more general covariant metric theories, the equations of motion typically become higher order in derivatives. Such higher order theories of gravity introduce a wide variety of interesting features at both classical and quantum level, but unfortunately one of them is a generic pathology in the guise of a ghost \cite{Stelle:1976gc,Capozziello:2011et,Clifton:2011jh}.

The ghosts may however be absent in a full nonperturbative theory \cite{Biswas:2011ar,Biswas:2013kla}. Besides avoiding pathologies of their finite-order derivative truncations, nonlocal theories of gravity can admit (super-)renormalisation \cite{Tomboulis:1997gg,Modesto:2011kw} and suggest possible resolution of singularities in black holes \cite{Bambi:2013gva,Frolov:2015bta} and cosmology \cite{Biswas:2010zk,Biswas:2012bp}. On the other hand, nonlocalities in the infrared end \cite{Mitsou:2015yfa,Conroy:2014eja}, that are expected in loop-corrected effective theories  \cite{Barvinsky:2014lja,Codello:2015mba}, may provide hints towards solutions to the problems of the cosmological constant \cite{ArkaniHamed:2002fu,Jaccard:2013gla}, dark energy \cite{Koivisto:2008xfa,Woodard:2014iga} or dark matter \cite{Hehl:2008eu,Deffayet:2014lba}. The most promising paths of generalising purely metric gravity thus seem to naturally lead to nonlocal theories. For recently proposed models of nonlocal gravity, see also e.g. \cite{Maggiore:2013mea,Modesto:2013jea,Dragovich:2015iza,Dimitrijevic:2015eaa}.

Other avenues can be pursued in non-Riemannian spacetimes, in which the metric is not the only gravitational field but the connection is regarded as an independent object. Such spacetimes may emerge from various subtly different theoretical postulates: one may straightforwardly promote the connection into an independent variable (the {\it Palatini variational principle} \cite{Hehl:1994ue,Olmo:2011uz}), possibly in the company of the metric Levi-Civita connection as well (the {\it hybrid metric-Palatini theories} \cite{Harko:2011nh,Capozziello:2012ny}), consider it to emerge from a independent tensor field (the {\it bimetric variational principle} \cite{BeltranJimenez:2012sz,Golovnev:2014aca}) or free only some degrees of freedom to propagate as in Weyl \cite{Jimenez:2014rna,Haghani:2014zra} or more general \cite{Heinicke:2005bp,Karahan:2011zq} {\it distorted geometry}
\cite{0264-9381-8-9-004,Jimenez:2015fva}, or finally, parameterise a curvature-dependent relation between the metric and the connection in order to describe generic theories in a unified manner  (the {\it C-theory} \cite{Amendola:2010bk,Koivisto:2011vq}). What is thus common to these approaches is the distinction between the metric and the affine structures of the geometry, but in most cases the latter can as well be at least a posteriori associated to a metric. 

Such theories featuring two metrics can have interesting relations to nonlocal gravity. In the case of a bimetric theory belonging to the ghost-free 4-parameter family of massive bimetric theories \cite{Hassan:2011tf}, it was shown explicitly at quadratic order in curvature that by integrating out one of the metrics one obtains an equivalent nonlocal formulation of the theory \cite{Cusin:2014zoa}. If one integrates out a metric using instead the equation of motion for the other metric, one can establish a correspondence, at the level of equations of motion, to fourth order conformal gravity \cite{Hassan:2013pca}, exhibiting a gauge symmetry which may furthermore be extended to higher, perhaps to all, orders \cite{Hassan:2015tba}. 

Here we will investigate nonlocal formulations of metric-affine gravity, to be specific, in the context of the C- and the so called D-theory. Their limiting Palatini theories can avoid introducing new propagating degrees of freedom in the pure gravity sector by adding no derivatives, and nonlocal theories by adding an infinite number of them, suggesting that the former could provide an effective description of the latter \cite{Koivisto:2009jn}. Some specific C-theory actions in fact have been shown to be classically nothing but nonlocal gravity involving inverse d'Alembertian operators acting upon the scalar curvature \cite{Sandstad:2013oja}. A question that was left open concerned tensorial nonlocalities. Such, unlike those constructed solely from the Ricci scalar invariant, modify the graviton propagator and could thus in principle adjust the theory beneficially by for example alleviating the ultraviolet divergences, but it is not clear whether such tensor nonlocalities in the gravity sector could emerge, in particular in their ghost-free form, as an effective description of generalised space-time structure. To investigate this, we need to consider the somewhat more general D-theory - that is, gravity in spacetime wherein the two fundamental geometrical structures are non-conformally related.

In the following we will first in Section \ref{nonp} set up the framework and derive a formal relation between C-theories and purely metric actions at the full nonperturbative level. In Section \ref{lin1} we restrict to quadratic order in metric fluctuations in order to obtain explicit results: we then derive two different effective formulations of the theory, and discuss their (in)equivalences at the level action and that of the equations of motion. In Section \ref{toy} we summarise the results and illustrate some basic lessons from the derivations by means of a simplified scalar toy model. 

\section{C-theory and nonlocal gravity}
\label{nonp}

In this Section we will first introduce the C-theory, establish its formal relation to purely metric gravity and briefly remark how the result (\ref{hataction}, \ref{rel1}) generalises a finding of Ref. \cite{Amendola:2010bk} and corrects another in Ref. \cite{Sandstad:2013oja}.
In part \ref{allord} we then give a general perturbative procedure to obtain the non-local Riemannian picture explicitly.

\subsection{C- and D-theories}

Let us consider an Einstein-Hilbert action
\be \label{action}
S=\int {\rm d}^nx \sqrt{-\rm{det}(g)} g^{\mu\nu}{\hat R}_{\mu\nu}\,,
\ee
that is generalised in such a way that the ''hatted'' Ricci tensor corresponds to a connection that is not simply the Levi-Civita connection as GR postulates. To be explicit, the curvature tensor is given in terms of the connection $\hat{\Gamma}^\al_{\bt\ga}$ by the usual formula
\be
{\hat R}_{\mu\nu} = 
\hat{\Gamma}^\alpha_{\mu\nu , \alpha}
       - \hat{\Gamma}^\alpha_{\mu\alpha , \nu} +
\hat{\Gamma}^\alpha_{\alpha\lambda}\hat{\Gamma}^\lambda_{\mu\nu} -
\hat{\Gamma}^\alpha_{\mu\lambda}\hat{\Gamma}^\lambda_{\alpha\nu}\,.\label{r_def} 
\ee
Throughout this paper we consider only symmetric, i.e. torsion-free connections, and only symmetric Ricci tensors.
Now, however, the connection is taken to be compatible with the metric $\hat{g}_{\al\bt}$ that is related to the physical metric $g_{\al\bt}$ as follows:
\be \label{relation}
{\hat g}_{\mu\nu}=C(\mathbf R)g_{\mu\nu}+D(\mathbf R){\hat R}_{\mu\nu}\,.
\ee
The functions $C$ and $D$ depend on the matrix ${\mathbf R}^{\mu}_{\nu}\equiv g^{\mu\alpha}{\hat R}_{\alpha\nu}$ via its scalar invariants such as $\mathrm{Tr}{\mathbf R}=g^{\mu\nu}{\hat R}_{\mu\nu}\equiv{\mathcal R}$, $\mathrm{Tr}{\mathbf R}^2$ et cetera, and their arbitrary derivatives such as $\square {\mathcal R}$, $\mathrm{Tr}[(\nabla\mathrm{Tr}{\nabla\mathbf R}){\mathbf R}]$ et cetera. Thus, for generality, we can allow any kind of covariant curvature dependence for the functions $C$ and $D$ that define the theory. The fundamental relation (\ref{relation}) is much more general than the prototype model \cite{Amendola:2010bk} with $C=C({\mathcal R})$, $D=0$ that suffices to cover the $f(R)$ -type models in their metric, Palatini and non-minimally coupled versions (and was used to study e.g. their cosmology \cite{Koivisto:2013xza} and Newtonian limit \cite{Koivisto:2011tp}).
As mentioned in the introduction however, an interest of ours is in tensorial nonlocalities due to their capability to modify the graviton propagator, and therefore we adopt the starting point (\ref{relation}) (though for some of the following computations we will include only nonzero $C$ for simplicity). 

As the metric $\hat{g}_{\mu\nu}$ defining the affine structure, and the metric $g_{\mu\nu}$ defining the geometry for matter fields, have a prescribed relation (\ref{relation}), it should be in principle possible to rewrite the theory in terms of just one effective metric. It turns out that this is more feasible in terms of the metric $\hat{g}_{\mu\nu}$. To this end, let us also define a matrix ${\mathbf{\hat R}}^{\mu}_{\nu}\equiv {\hat g}^{\mu\alpha}{\hat R}_{\alpha\nu}$. We have then:
\be
g_{\mu\nu}=\frac{{\hat g}_{\mu\alpha}}{C}\left(\delta^{\alpha}_{\nu}-D{\hat R}^{\alpha}_{\nu}\right)\,,
\ee
or, in matrix notation,
\be
g=\frac1C\cdot {\hat g}\left(I-D{\mathbf{\hat R}}\right)\,,
\ee
and for the inverse
\be \label{inverse}
g^{-1}=C\cdot\left(I-D{\mathbf{\hat R}}\right)^{-1}{\hat g}^{-1}\,.
\ee
We then find that the action reads
\be \label{hataction}
S=\int {\rm d}^nx\sqrt{-\rm{det} \hat g}\cdot \frac{\sqrt{\rm{det} \left(I-D{\mathbf{\hat R}}\right)}}{C^{\frac{n-2}{2}}}\cdot \rm{Tr}\left(\left(I-D{\mathbf{\hat R}}\right)^{-1}{\mathbf{\hat R}}\right)\,.
\ee
In order to have the theory solely in terms of the metric $\hat{g}_{\mu\nu}$, one has to rephrase the arguments of $C$ and $D$ functions in terms of hatted quantities. For that we multiply the equation for $g^{-1}$ by ${\hat R}_{\mu\nu}$ and get an equation for $\mathbf R$ in terms of $\mathbf{\hat R}$:
\be \label{rel1}
\mathbf R=C(\mathbf R)\cdot\left(I-D(\mathbf R)\cdot{\mathbf{\hat R}}\right)^{-1}\mathbf{\hat R}\,.
\ee
This equation can be solved - at least in principle - for the  $\mathbf R$. With the result plugged back into (\ref{hataction}), we've arrived at the goal.

This generalises the formulation of these theories in the  ''$C$-frame'' as put forward in Ref. \cite{Amendola:2010bk}. In the particular case of $C=C(\mathcal{R})$, $D=0$, the explicit formulation becomes very simple. Assume for example $C \sim \mathcal{R}^\alpha$. We then obtain from (\ref{rel1}) that $\mathcal{R} \sim \hat{R}^{\frac{1}{1-\alpha}}$. Thus the action
(\ref{hataction}) becomes that of a power-law $f(\hat{R})$ model, in particular $f \sim \hat{R}^{\frac{1-n\alpha/2}{1-\alpha}}$, in accordance with the result of \cite{Amendola:2010bk}.

Note that there is an erroneous statement in \cite{Sandstad:2013oja} that the model is fully equivalent to pure GR for $C \sim \left( 1+A\mathcal{R}\right)^{\frac{4}{n-2}}$, $D=0$. We see from (\ref{hataction}) and (\ref{rel1}) that it is instead of a non-trivial $f(R)$ type. The source of the mistake is that the coefficient $\beta$ in their formula (9) is curvature-dependent while only its value $\frac{4A(n-1)}{n-2}$ at zero curvature is taken in calculations of the Ref.  \cite{Sandstad:2013oja}.

%In the presence of matter, the model is not simply an $f(R)$ model but comes also with a non-minimal coupling that violates the covariant energy-momentum conservation \cite{Koivisto:2005yk, Harko:2014gwa}. For this reason the $C$-theory framework can be used to study wide a range of models, covering also the non-minimally coupled models,  in a unified manner, as demonstrated in a PPN computation in Ref. \cite{Koivisto:2011tp}. The ''$C$-frame'' formulation of the teories with completely general curvature dependence, given by (\ref{hataction}) together with (\ref{rel1}) extends this framework to models with arbitrary curvature couplings. 

%However, we expect that only specific cases of such models might be viable in terms of avoiding ghosts. In the following we will make some steps towards identifying these potentially viable classes of models. It turns out that much can be uncovered already by considering these theories at the linear level.

\subsection{Calculation of ${\mathcal R}$ to all orders in non-local picture} 
\label{allord}

Let us solve the curvature relation for a C-model ($D=0$) with 
\be
C({\mathcal R})\equiv 1+\sum_{i=1}^{\infty}c_{,i}{\mathcal R}^i\,,
\ee
to all orders in curvature. It is easy to see  \cite{Sandstad:2013oja} that
\begin{equation}
\label{currel}
{\mathcal R}=R-(n-1)\frac{\square C}{C}-(n-1)(n-6)\frac{(\partial C)^2}{4C^2}\,.
\end{equation}

Obviously, to the first order we have %${\mathcal R}\approx R-(n-1)\square c_{,1}{\mathcal R}$, and thus
\be
{\mathcal R}\approx R-(n-1)\square c_{,1}{\mathcal R}\,,
\ee
and
\be
\label{fiorR}
{\mathcal R}\approx \frac{1}{1+(n-1)\square c_{,1}}R\,,
\ee
where the fraction should be understood as a non-local operator acting on $R$ (and we take into account that $\square$ does not necessarily commute with $c_{,1}$). Of course, the usual issues with the precise definition of non-local operators are there. Concerning this problem, we would not go beyond the standard treatments in this paper.

Now we want to go to the second order and for this purpose put
\be
{\mathcal R}=\frac{1}{1+(n-1)\square c_{,1}}R+X_2
\ee
into the curvature relation (\ref{currel}), where $X_2$ then represents the second order correction. At this order, we obtain:
\begin{eqnarray}
\label{secondcur}
X_2 & = & \frac{1}{1+(n-1)\square c_{,1}}\left[ (n-1)\left(c_{,1}\frac{1}{1+(n-1)\square c_{,1}}R\right)\square \left(c_{,1}\frac{1}{1+(n-1)\square c_{,1}}R\right)\right. \nonumber \\ & - & \left.\frac{(n-1)(n-6)}{4}\left(\partial_{\mu}\left(c_{,1}\frac{1}{1+(n-1)\square c_{,1}}R\right)\right)^2\right] 
-(n-1)\square c_{,2}\left(\frac{1}{1+(n-1)\square c_{,1}}R\right)^2\,.
\end{eqnarray}

One can easily see that the general structure of the higher orders will be the same:
\be
X_n=\frac{1}{1+(n-1)\square c_{,1}}\cdot {\mathfrak F}\left(X_i,\ \square X_i, \ (\partial_{\mu}X_i)^2\right) - (n-1)\square c_{,n}\left(\mathcal{R}^{1}\right)^n\,,
\ee
with $i=1,\ldots, n-1$, and $\mathcal{R}^{1}$ is the first order expression for $\mathcal{R}$ obtained above (\ref{fiorR}). Therefore, up to the precise definition for the operator $(1+(n-1)\square c_{,1})^{-1}$, one gets a full series expansion for ${\mathcal R}$ as a function of $R$ .

If one is allowed to integrate by parts in the second to last term in eq. (\ref{secondcur}) and ignore the final total derivative term with $c_{,2}$, then the second order action is
\be
\label{sord}
S=\int {\rm d}^4x  \sqrt{-\rm{det}(g)}\cdot\frac{1}{1+(n-1)\square c_{,1}}\left(R+\frac{(n-1)(n-2)}{4}\left(c_{,1}\frac{1}{1+(n-1)\square c_{,1}}R\right)\square \left(c_{,1}\frac{1}{1+(n-1)\square c_{,1}}R\right)\right)\,,
\ee
and if $c_{,1}$ is a number then the first non-local operator might be regarded as a unity plus total derivative terms. In the following we will study this theory in more detail.

\section{Linearised theory}
\label{lin1}

In this Section we consider the theory (\ref{action},\ref{relation}) at the quadratic order in perturbations. We will rewrite the theory at this order first in terms of the spacetime metric $\hat{g}_{\mu\nu}$ in \ref{lhat} and then in terms of the physical metric $g_{\mu\nu}$ in \ref{lnonhat}. Under some restrictions the resulting formulations should be equivalent at the level of equations of motions, which we verify in \ref{eom}.

\subsection{Local $\hat g$-formulation}
\label{lhat}

Let us consider perturbatively the action (\ref{action}) where the curvature is implictly given by the relation (\ref{relation}). 
We study the perturbations around the double Minkowski solution and denote at ${\hat R}_{\mu\nu}\to0$
\be
C(\mathbf R)=1+c_{,1}\mathcal R+ \cdots\,, \quad \quad 
D(\mathbf R)=d_0+ \cdots\,.
\ee
We have for first-order fluctuations ($g\equiv\eta+h$):
\be
\label{linmetrrel}
{\hat h}_{\mu\nu}=h_{\mu\nu}+c_{,1}\mathcal R \eta_{\mu\nu}+d_0 {\hat R}_{\mu\nu}\,,
\ee
and for curvatures:
\be
{\hat R}_{\mu\nu}=\frac12\left(\partial^2_{\mu\alpha}{\hat h}^{\alpha}_{\nu}+\partial^2_{\nu\alpha}{\hat h}^{\alpha}_{\mu}-\square {\hat h}_{\mu\nu}-\partial^2_{\mu\nu}{\hat h}^{\alpha}_{\alpha}\right)+ \cdots
\ee
and
\be
\mathcal R=\partial^2_{\mu\nu}{\hat h}^{\mu\nu}-\square {\hat h}^{\mu}_{\mu}+ \cdots
\ee
where the indices are raised with $\eta_{\mu\nu}$.
The second order action reads
\be
S=\int {\rm d}^nx\left(\eta^{\mu\nu}\delta^{(2)}{\hat R}_{\mu\nu}+\delta^{(1)}\left(\sqrt{-\rm{det}(\hat{g})}g^{\mu\nu}\right)\cdot\delta^{(1)}{\hat R}_{\mu\nu}\right)\,.
\ee
where $\delta^{(n)}(X)$ is the $n$th order contribution to the operator $X$.
Using
\be
\delta^{(1)}\left(\sqrt{-\rm{det}({g})}g^{\mu\nu}\right)=-h^{\mu\nu}+\frac12 \eta^{\mu\nu}h^{\alpha}_{\alpha}\,,
\ee
and
\be
\delta^{(1)}\left(\sqrt{-\rm{det}(\hat{g})}{\hat g}^{\mu\nu}\right)=\delta^{(1)}\left(\sqrt{-g}g^{\mu\nu}\right)-\frac12\eta^{\mu\nu}\left((n-2)c_{,1}+d_0\right){\hat R}+d_0{\hat R}^{\mu\nu}\,,
\ee
and taking into account that
\be
\eta^{\mu\nu}\delta^{(2)}{\hat R}_{\mu\nu}+\delta^{(1)}\left(\sqrt{-\rm{det}(\hat{g})}{\hat g}^{\mu\nu}\right)\cdot\delta^{(1)}{\hat R}_{\mu\nu}=\delta^{(2)}\left(\sqrt{-\rm{det}(\hat{g})}{\hat R}\right)\,,
\ee
we see that the model is equivalent at quadratic level to
\be \label{pert}
S=\int {\rm d}^nx \sqrt{-\rm{det}(\hat{g})}\left(\hat R-\frac{(n-2)c_{,1}+d_0}{2}{\hat R}^2+d_0 {\hat R}^{\mu\nu}{\hat R}_{\mu\nu}\right)\,,
\ee
and therefore contains ghosts unless it is a pure C-theory, i.e. $D=0$. However, this could be avoided if the relation (\ref{relation}) is given by nonlocal functions $C$ and $D$, as we will show next.

\subsubsection{The case of non-local relations $C$, $D$}

We can allow the coefficients $c_{,i} \equiv c_{,i}(\square)$ and $d_{,i}\equiv d_{,i}(\square)$ to be arbitrary functions of the covariant d'Alembertian operator
$\square \equiv g^{\mu\nu}\nabla_\mu\nabla_\nu$. A brief re-examination of the derivation of eq. (\ref{pert}) with some care on the differential operator ordering shows that at the quadratic order the theory can be written as
\be \label{pert-nl}
S=\int {\rm d}^nx \sqrt{-\rm{det}(\hat{g})}\left(\hat R- {\hat R}\frac{(n-2)c_{,1}(\square)+d_0(\square)}{2}{\hat R}+{\hat R}_{\mu\nu}d_0(\square) {\hat R}^{\mu\nu}\right)\,.
\ee
Using the formulae presented in Refs. \cite{Biswas:2011ar,Biswas:2013kla}, we can directly write the propagator corresponding to any metric action. The propagator $\Pi_{\al\bt\gamma\delta}$, which in vacuum satisfies $\Pi^{-1}_{\al\bt\gamma\delta}h^{\al\bt}=0$, is given for the action (\ref{pert-nl}) in flat 4-dimensional ($n=4$) Fourier space ($\square \rightarrow -k^2$) by
\be
k^2\Pi = \frac{\Pi^{(2)}}{1+\frac{1}{2}d_0k^2} - \frac{\Pi^{(0)}}{2+(6c_{,1}+d_0)k^2}\,,
\ee
where $\Pi^{(2)}$ is a spin-2 projector and $\Pi^{(0)}$ is a spin-0 projector. We refer the reader to Refs. \cite{Biswas:2011ar,Biswas:2013kla} for a detailed exposition of the formalism, but here it  suffices to notice that in GR we have that $\Pi_{GR} \sim \Pi^{(2)}-\Pi^{(0)}/2$. Therefore, by making the choice
\be \label{d0}
d_0 = -3c_{,1}+2/\square\,, 
\ee
we can avoid introducing extra degrees of freedom. This occurs because the propagator then becomes pure a modulation of the usual GR propagator,
\be
\Pi = \frac{2\Pi_{GR}}{3c_{,1}k^2}\,.
\ee
Now, the equation $c_{,1}(-k^2)k^2=0$ should not have solutions, since they would introduce additional poles in the above propagator, and these poles would always represent ghost-like tensor modes. (Of course, positive powers of $k^2$ would tend to introduce infrared problems anyway, but we would not consider graviton scattering here and prefer to state these requirements explicitly.) However, by considering the class of relations (\ref{d0}) we restrict to just modifying the propagation of the GR graviton mode. As an example, consider the function $c_{,1} = M^2 e^{\square/M^2}/\square$ where $M$ is an ultraviolet mass scale. This results in precisely the kind of exponential nonlocality that was proposed in Ref. \cite{Biswas:2011ar} as a means of taming the ultraviolet divergences of GR and yielding a ghost and singularity free theory of gravity. We have thus seen that such theories could be interpreted as a manifestation of the nonlocal relation between the spacetime connection and the metric a'la $C$-theories.

\subsection{Non-local $g$-formulation for C-models}
\label{lnonhat}

Above we eliminated the physical metric $g_{\mu\nu}$ and wrote the resulting quadratic action (\ref{pert}) in terms of the other metric. This turns out to be easier in practice than eliminating 
 $\hat g$  in order to get a (possibly non-local) action for $g$. However, as we have seen, the latter can also be done at least for C-theories at perturbative level around double Minkowski. 

Let us first naively start with the linear level relation (\ref{linmetrrel}) between the metrics.  If we for convenience write the general Weyl relation as  
${\hat g}_{\mu\nu}=e^{2\rho}g_{\mu\nu}$, eq. (\ref{currel}) with $C=e^{2\rho}$ gives ${\mathcal R}=R-(n-1)(n-2)(\partial\rho)^2-2(n-1)\square\rho$.
It is not difficult to find the $\rho$-factor to the first order in perturbations. To this end, we first write
\be\label{h-rel}
{\hat h}_{\mu\nu}=h_{\mu\nu}+c_{,1}\eta_{\mu\nu}\left(\partial^2_{\alpha\beta}{\hat h}^{\alpha\beta}-\square {\hat h}^{\alpha}_{\alpha}\right)\,,
\ee
and then note that only the trace part is changed,
\be
{\hat h}_{\mu\nu}=h_{\mu\nu}+\frac1n \eta_{\mu\nu}\left( {\hat h}^{\alpha}_{\alpha}-h^{\alpha}_{\alpha}\right)\,,
\ee
and that it thus can be solved as
\be\label{mrnl}
{\hat h}^{\mu}_{\mu}=\frac{1}{1+(n-1)c_{,1}\square}\left( h^{\mu}_{\mu}+nc_{,1}\partial^2_{\mu\nu}h^{\mu\nu}-c_{,1}\square h^{\alpha}_{\alpha}\right)\,,
\ee
which with this accuracy is equivalent to Weyl transformation with
\be
\rho=\frac{1}{2n}({\hat h}^{\alpha}_{\alpha}-h^{\alpha}_{\alpha})=\frac{c_{,1}\left(\partial^2_{\alpha\beta}{h}^{\alpha\beta}-\square {h}^{\alpha}_{\alpha}\right)}{2(1+(n-1)c_{,1}\square)}=\frac{c_{,1}}{2(1+(n-1)c_{,1}\square)}R\,.
\ee
Therefore the action is
\be 
S=\int {\rm d}^nx\sqrt{-\rm{det}({g})}\left(R-(n-1)(n-2)(\partial\rho)^2-2(n-1)\square\rho\right)\,,
\ee
or, using partial integration and keeping all surface terms, the explicit action becomes:
\be 
S=\int {\rm d}^nx\sqrt{-\rm{det}({g})}\left(R+R\frac{(n-1)(n-2)c^2_{,1}\square}{4(1+(n-1)c_{,1}\square)^2}R-\partial_{\mu}\left(R\frac{(n-1)(n-2)c^2_{,1}\partial^{\mu}}{4(1+(n-1)c_{,1}\square)^2}R\right)-\frac{(n-1)c_{,1}\square}{1+(n-1)c_{,1}\square}R\right),
\ee
and, if the surface terms can be omitted (otherwise we must use the C-relation to the second order including the $c_{,2}$ part), we get
\be \label{nonlocal}
S=\int {\rm d}^nx\sqrt{-\rm{det}({g})}\left(R-(n-1)(n-2)(\partial\rho)^2\right)=\int {\rm d}^nx\sqrt{-\rm{det}(g)}\left(R+R\frac{(n-1)(n-2)c^2_{,1}\square}{4(1+(n-1)c_{,1}\square)^2}R\right)\,.
\ee
Modulo surface terms, this action is equivalent to (\ref{sord}) if the operator $\frac{1}{1 + (n-1)c_{,1}\square}=\sum\limits_{n=0}^{\infty}\left(\vphantom{\sum}-(n-1)c_{,1}\square\right)^n$ can be treated as unity plus surface terms which we omit (for full equivalence it would have been necessary to use the second order accuracy in the conformal factor above). Of course, omitting such a non-local factor is not innocuous as we shall see later.

In terms of the physical metric $g_{\mu\nu}$, the C-theory can thus be effectively described as infrared nonlocally modified gravity, even when the function $C$ does not involve any additional derivative operators. The action (\ref{nonlocal}) resembles the $Rf(R/\square)$ models (and can be brought precisely into that form by a suitable choice of $c_{,1}$) that have been studied extensively in recent years, see for example \cite{Deser:2007jk,Koivisto:2008xfa,Elizalde:2013dlt}. Taking into account the finite $c_{,1}$, the inverse-d'Alembertian operator is regulated along the lines already discussed in Ref.\cite{Wetterich:1997bz}. 

The result (\ref{nonlocal}) agrees with the corresponding limit we arrived at in \ref{allord}. A more nontrivial consistency check is provided by checking the relation to the alternative $\hat{g}$-formulation of the theory we derived in the previous subsection. Before turning to this, we will however take a more careful look at some subtleties in these derivations in the case of nontrivial fundamental relation (\ref{relation}). We can already note though that, at the level of full Lagrangian densities, equivalence to the quadratic model from the previous section can be proven immediately. Indeed, in 
\be
S=\int {\rm d}^nx\sqrt{-\rm{det}({g})}\left(R-(n-1)(n-2)(\partial\rho)^2-2(n-1)\square\rho\right)=\int {\rm d}^nx\sqrt{-\rm{det}({g})}g^{\mu\nu}{\hat R}_{\mu\nu}\,,
\ee
we can make the conformal rescaling from $g$ to $\hat g$:
$$\int {\rm d}^nx\sqrt{-\rm{det}({\hat{g}})}\left(e^{-2\rho}\right)^{\frac{n-2}{2}}{\hat g}^{\mu\nu}{\hat R}_{\mu\nu}$$
and set $\left(e^{-2\rho}\right)^{\frac{n-2}{2}}\approx 1-\frac{n-2}{2}\cdot2\rho$ and $2\rho\approx c_{,1}{\mathcal R}\approx c_{,1}{\hat R}$, so that we get precisely the equation (\ref{pert}).

\subsubsection{The case of non-local relation $C$}

One has to be careful about the class of variations and boundary terms especially when the coefficients $c_{,i}$ are promoted into (inverse) derivative operators. For example, if we take a non-local function with $c_{,1}\sim\frac{1}{\square}$, then omitting the $\square\rho$-term does not seem even naively appropriate. In this case one might argue that we need to know $\rho$ to the second order in perturbations including the knowledge of $c_{,2}$. 

%However, at the level of full Lagrangian densities, equivalence to the quadratic model from the previous section can be proven. Indeed, in 
%\be
%S=\int d^nx\sqrt{-g}\left(R-(n-1)(n-2)(\partial\rho)^2-2(n-1)\square\rho\right)=\int d^nx\sqrt{-g}g^{\mu\nu}{\hat R}_{\mu\nu}
%\ee
%we can make the conformal rescaling from $g$ to $\hat g$:
%$$\int d^nx\sqrt{-\hat g}\left(e^{-2\rho}\right)^{\frac{n-2}{2}}{\hat g}^{\mu\nu}{\hat R}_{\mu\nu}$$
%and setting $\left(e^{-2\rho}\right)^{\frac{n-2}{2}}\approx 1-\frac{n-2}{2}\cdot2\rho$ and $2\rho\approx c_{,1}{\mathcal R}\approx c_{,1}{\hat R}$, we get precisely the equation (\ref{pert}). 

%Barring these complications, we see that, in terms of the physical metric $g_{\mu\nu}$, the C-theories thus are naturally interpreted as infrared nonlocally modified gravity, even when the function $C$ does not involve any additional derivative operators. The action (\ref{nonlocal}) resembles the $Rf(R/\square)$ models (and can be brought precisely into that form by a suitable choice of $c_{,1}$) that have been studied extensively in recent years, see for example \cite{Deser:2007jk,Koivisto:2008xfa,Elizalde:2013dlt}. Taking into account the finite $c_{,1}$ the inverse-d'Alembertian operator is regulated along the lines already discussed in Ref.\cite{Wetterich:1997bz}. 

%Strictly speaking, we need to be very accurate about the boundary terms. 
The action (\ref{action}) contains the linear (surface) term $\partial^2_{\mu\nu}{\hat h}^{\mu\nu}-\square {\hat h}^{\mu}_{\mu}$ which requires the second order accuracy in the relation between the metrics (\ref{relation}). Let us now show how to perform an accurate quadratic level treatment of the action (\ref{action}) in terms of metric fluctuations. For the sake of completeness, we give the second order expressions for all the relevant quantities (all indices are raised with the background Minkowski metric): \newline
the inverse metric
\be
\label{metric2ord}
g^{\mu\nu}=\eta^{\mu\nu}-h^{\mu\nu}+h^{\mu\alpha}h^{\nu}_{\alpha}+{\mathcal O}\left(h^3\right)\,,
\ee
the metric determinant
\be
\label{sqrt2ord}
\sqrt{-\rm{det}({g})}=1+\frac12 h^{\mu}_{\mu}-\frac14 h_{\mu\nu}h^{\mu\nu}+\frac18 \left(h^{\mu}_{\mu}\right)^2+{\mathcal O}\left(h^3\right),
\ee
connection coefficients
\be
\label{Gamma2ord}
{\hat \Gamma}^{\alpha}_{\mu\nu}=\frac12 \left(\partial_{\mu}{\hat h}^{\alpha}_{\nu}+\partial_{\nu}{\hat h}^{\alpha}_{\mu}-\partial^{\alpha}{\hat h}_{\mu\nu}\right)-\frac12 {\hat h}^{\alpha\beta}\left(\partial_{\mu}{\hat h}_{\beta\nu}+\partial_{\nu}{\hat h}_{\beta\mu}-\partial_{\beta}{\hat h}_{\mu\nu}\right)+{\mathcal O}\left({\hat h}^3\right),
\ee
the mixed curvature invariant
\begin{multline}
\label{curlyR2ord}
{\mathcal R}\equiv g^{\mu\nu}{\hat R}_{\mu\nu}=
\partial^2_{\mu\nu}{\hat h}^{\mu\nu}-\square {\hat h}^{\mu}_{\mu}-\frac12 h^{\mu\nu}\left(\partial^2_{\mu\alpha}{\hat h}^{\alpha}_{\nu}+\partial^2_{\nu\alpha}{\hat h}^{\alpha}_{\mu}-\square{\hat h}_{\mu\nu}-\partial^2_{\mu\nu}{\hat h}^{\alpha}_{\alpha}\right)\\
+\frac14(\partial_{\mu}{\hat h}_{\alpha\beta})(\partial^{\mu}{\hat h}^{\alpha\beta})-\frac12 (\partial_{\mu}{\hat h}_{\alpha\beta})(\partial^{\alpha}{\hat h}^{\mu\beta})+\frac12(\partial_{\mu}{\hat h}^{\mu\alpha})(\partial_{\alpha}{\hat h}^{\beta}_{\beta})-\frac14(\partial_{\mu}{\hat h}^{\alpha}_{\alpha})(\partial^{\mu}{\hat h}^{\beta}_{\beta})\\
-\partial_{\alpha}\left({\hat h}^{\alpha\beta}\left(\partial_{\mu}{\hat h}^{\mu}_{\beta}-\frac12 \partial_{\beta}{\hat h}^{\mu}_{\mu}\right)\right)+\frac12\partial^{\mu}\left({\hat h}^{\alpha\beta}\partial_{\mu}{\hat h}_{\alpha\beta}\right)
+{\mathcal O}\left(\left(h,{\hat h}\right)^3\right),
\end{multline}
and, after some elementary rearrangements, the Lagrangian density
\begin{multline}
\label{lagrangian2ord}
\sqrt{-\rm{det}({g})}{\mathcal R}\approx\left(1+\frac12 h^{\alpha}_{\alpha}\right){\mathcal R}=-\frac12(\partial_{\mu}{\hat h}_{\alpha\beta})\left(\partial^{\mu}h^{\alpha\beta}-\frac12 \partial^{\mu}{\hat h}^{\alpha\beta}\right)+(\partial_{\mu}{\hat h}_{\alpha\beta})\left(\partial^{\alpha}h^{\mu\beta}-\frac12\partial^{\alpha}{\hat h}^{\mu\beta}\right)\\
-\frac12\left((\partial_{\mu}h^{\mu\alpha})(\partial_{\alpha}{\hat h}^{\beta}_{\beta})+(\partial_{\mu}{\hat h}^{\mu\alpha})(\partial_{\alpha}h^{\beta}_{\beta})-(\partial_{\mu}{\hat h}^{\mu\alpha})(\partial_{\alpha}{\hat h}^{\beta}_{\beta})\right)+\frac12(\partial_{\mu}{\hat h}^{\alpha}_{\alpha})\left(\partial^{\mu}h^{\beta}_{\beta}-\frac12\partial^{\mu}{\hat h}^{\beta}_{\beta}\right)\\
-\partial_{\alpha}\left({\hat h}^{\alpha\beta}\partial_{\mu}{\hat h}^{\mu}_{\beta}+h^{\mu\beta}\partial_{\mu}{\hat h}^{\alpha}_{\beta}-\frac12 \left({\hat h}^{\alpha\beta}+h^{\alpha\beta}\right)\partial_{\beta}{\hat h}^{\mu}_{\mu}\right)+\frac12\partial^{\mu}\left(\left({\hat h}^{\alpha\beta}+h^{\alpha\beta}\right)\partial_{\mu}{\hat h}_{\alpha\beta}\right)\\
+\frac12 \partial_{\mu}\left(h^{\alpha}_{\alpha}\left(\partial_{\nu}{\hat h}^{\nu\mu}-\partial^{\mu}{\hat h}^{\nu}_{\nu}\right)\right)
+\partial^2_{\mu\nu}{\hat h}^{\mu\nu}-\square {\hat h}^{\mu}_{\mu}+{\mathcal O}\left(\left(h,{\hat h}\right)^3\right).
\end{multline}
It is easy to see that for $h=\hat h$ it gives the standard quadratic GR.

Now we see that, indeed, in order to go to the $\hat g$ picture we need the relation between $h$ and $\hat h$ only to the first order since $h$ (which we want to exclude) enters only in quadratic terms. However, transition to the picture of $g$ requires the second order accuracy for $\hat h$ in terms of $h$ if we are to keep proper track of the surface term $\partial^2_{\mu\nu}{\hat h}^{\mu\nu}-\square {\hat h}^{\mu}_{\mu}$ in the action. The relation (\ref{relation}) then takes the form
\be
{\hat h}_{\mu\nu}=\left(1+c_{,1}\delta{\mathcal R}^{(1)}\right)h_{\mu\nu}+\left(c_{,1}\delta{\mathcal R}^{(2)}+\frac12 c_{,2}\left(\delta{\mathcal R}^{(1)}\right)^2\right)\eta_{\mu\nu}+{\mathcal O}\left(\left(h,{\hat h}\right)^3\right).
\ee
One can solve this relation to second order using the results of the Section \ref{allord}. However, it would be very cumbersome. From the viewpoint of the $\hat h$ picture, we are doing a $c_{,2}$ -dependent change of variables to $h$. Therefore there is no direct contradiction in obtaining a $c_{,2}$ dependent model, though the $\hat{g}$-picture of \ref{lhat} was independent of the $c_{,2}$ up to the quadratic order in perturbations.

Relation (\ref{secondcur}) shows that the $c_{,2}$ dependence appears only via a surface term if $c_{,i}$ are ordinary functions and natural boundary conditions are used. To what extent the model with these boundary conditions is equivalent to the $\hat h$ picture will be explained below. In principle, derivative and/or non-local changes of variables might interfere with the boundary conditions. Note also that if $c_{,1}\propto c_{,2}\propto\frac{1}{\square}$ then, naively, the $c_{,2}$-term in the action is no longer a surface term. However, it is very important that the linear part of the action must be a surface term in any picture. Otherwise the double-Minkowski will no longer be a solution. If we are allowed to drop the surface terms, then this is the case for (\ref{lagrangian2ord}) after substituting (\ref{mrnl}).

\subsection{Equations of motion} 
\label{eom}

Let us illustrate an apparent discrepancy between the two pictures, given in \ref{lhat} and \ref{lnonhat}, respectively, by a specific example. A suitable special case is given in $n=4$ by choosing $c_{,1}=\frac{2}{3\square}$ and $D=0$ for simplicity. We see that $\frac{1}{1+(n-1)\square c_{,1}}=\frac{1}{3}$ and the 
$g$-picture action (\ref{nonlocal}) reduces to
\be 
S \approx \int {\rm d}^4x \sqrt{-\rm{det}({g})}\left(\frac13 R+\frac{2}{81}R\frac{1}{\square}R\right)\,,
\ee
which resembles the structure in the $\hat g$ picture (\ref{pert-nl}) that now reduces to
\be \label{pert-nl2}
%S=\int {\rm d}^nx \sqrt{-\rm{det}(\hat{g})}\left(\hat R- {\hat R}\frac{2c_{,1}(\square)}{2}{\hat R}\right) 
S \approx 
\int {\rm d}^nx \sqrt{-\rm{det}(\hat{g})}\left({\hat R}-\frac{2}{3}{\hat R}\frac{1}{\square}{\hat R}\right)\,.
\ee
however the coefficients and the physical spectrum are different.
Nevertheless, now we can explicitly check that, modulo the integration by parts (and higher order terms), the actions are equal. Indeed, with ${\mathcal R}=\frac13 R+\frac{2}{81}R\frac{1}{\square}R$ we have
\be
{\hat R}=\frac{{\mathcal R}}{C({\mathcal R})}={\mathcal R}-{\mathcal R}c_{,1}{\mathcal R}=\frac13 R-\frac{4}{81}R\frac{1}{\square}R\,,
\ee
and then $\sqrt{-\hat g}=C^2({\mathcal R})\sqrt{-g}=\left(1+\frac{4}{9\square}R\right)\sqrt{-g}$ to the linear order, and for the hatted action we obtain
\be
\sqrt{-\rm{det}(\hat{g})}\left({\hat R}-\frac{2}{3}{\hat R}\frac{1}{\square}{\hat R}\right)=\sqrt{-\rm{det}({g})}\left(\frac13 R+\frac{2}{81}R\frac{1}{\square}R\right)\,,
\ee
which proves the mathematical equality. However, even the signs of the correction terms differ in the two pictures, and it appears nontrivial that they ought to describe the same physics.

This can be clarified by checking the correspondence between different frames at the level of equations of motion, as we shall do in the following without fixing any of the coefficients $c_{,i}$.

One can easily find the equations of motion for the action (\ref{pert}):
\be
\left(1-(n-2)c_{,1}{\hat R}\right){\hat R}_{\mu\nu}+(n-2)c_{,1}\left({\hat\bigtriangledown}_{\mu}{\hat\bigtriangledown}_{\nu}-{\hat g}_{\mu\nu}{\hat\square}\right){\hat R}-\frac12\left(\hat R-\frac{(n-2)c_{,1}}{2}{\hat R}^2\right){\hat g}_{\mu\nu}=0\,,
\ee
which at the linear level boil down to
\be
{\hat R}_{\mu\nu}-\frac12 {\hat R}\eta_{\mu\nu}+(n-2)c_{,1}\left(\partial_{\mu}\partial_{\nu}-\eta_{\mu\nu}\square\right){\hat R}=0\,,
\ee
or, in explicit metric variables:
\be\label{lineqhat}
\partial^2_{\mu\alpha}{\hat h}^{\alpha}_{\nu}+\partial^2_{\nu\alpha}{\hat h}^{\alpha}_{\mu}-\square {\hat h}_{\mu\nu}-\partial^2_{\mu\nu}{\hat h}^{\alpha}_{\alpha}+\left(\vphantom{\int}2(n-2)c_{,1}\left(\partial^2_{\mu\nu}-\eta_{\mu\nu}\square\right)-\eta_{\mu\nu}\right)\cdot\left(\partial^2_{\alpha\beta}{\hat h}^{\alpha\beta}-\square{\hat h}^{\alpha}_{\alpha}\right)=0\,.
\ee

Let us choose the harmonic gauge for $\hat h$:
\be\label{harmonic}
\partial_{\mu}{\hat h}^{\mu\nu}=\frac12 \partial^{\nu}{\hat h}^{\mu}_{\mu}\,.
\ee
Then it is easy to see from (\ref{h-rel}) that
\be
{\hat h}_{\mu\nu}=h_{\mu\nu}-\frac{1}{2}c_{,1}\eta_{\mu\nu}\square {\hat h}^{\alpha}_{\alpha}\,,
\ee
or
\be
{\hat h}^{\mu}_{\mu}=\frac{1}{1+\frac{n}{2}c_{,1}\square}\cdot h^{\mu}_{\mu}\,,
\ee
and, in the picture without the hats, we have the gauge
\be\label{unhatgauge}
\partial_{\mu}h^{\mu\nu}=\frac12 \partial^{\nu}\frac{1+c_{,1}\square}{1+\frac{n}{2}c_{,1}\square} h^{\mu}_{\mu}\,.
\ee

Let us now apply the harmonic gauge (\ref{harmonic}) to the field equation  (\ref{lineqhat}):
\be
\square {\hat h}_{\mu\nu}+\frac12 \left(\vphantom{\int}2(n-2)c_{,1}\left(\partial^2_{\mu\nu}-\eta_{\mu\nu}\square\right)-\eta_{\mu\nu}\right)\cdot\square{\hat h}^{\alpha}_{\alpha}=0\,.
\ee
The traceless part obeys the wave equation with a source term dependent on the trace part, and for the trace part we have:
\be
(n-2)\left(\vphantom{\int}1+2(n-1)c_{,1}\square\right)\cdot\square{\hat h}^{\mu}_{\mu}=0\,,
\ee
which is equivalent to
\be \label{eomnonhat}
(n-2)\frac{1+2(n-1)c_{,1}\square}{1+\frac{n}{2}c_{,1}\square}\cdot\square h^{\mu}_{\mu}=0\,.
\ee

The question is now whether we get it also directly from the non-local action (\ref{nonlocal})? 
%The question is now whether we get it also directly from the non-local action (\ref{sord})? 

We find that the equation of motion to first order resulting from the action (\ref{nonlocal}) is\footnote{A comprehensive review on how to variate non-local actions to all orders can be found in \cite{Dimitrijevic:2015eza}.}:
\be
R_{\mu\nu}-\frac12 R\eta_{\mu\nu}-\frac{(n-1)(n-2)c_{,1}^2\square}{2\left(1 + (n-1)c_{,1}\square\right)^2}\left(\partial_{\mu}\partial_{\nu}-\eta_{\mu\nu}\square\right) R=0\,,
\ee
which in metric variables is:
\begin{eqnarray}
\partial_{\mu\alpha}^2h^\alpha_\nu + \partial_{\nu\alpha}^2h^\alpha_\mu - \square h_{\mu\nu} - \partial_{\mu\nu}^2h^\alpha_\alpha &&\nonumber\\
 - \left[\eta_{\mu\nu} + \frac{(n-1)(n-2)c_{,1}^2\square}{\left(1 + (n-1)c_{,1}\square\right)^2}\left(\partial_{\mu}\partial_{\nu}-\eta_{\mu\nu}\square\right)\right]\left(\partial_{\alpha\beta}h^{\alpha\beta} - \square h^{\alpha}_{\alpha}\right)&=&0\,,
\end{eqnarray}
employing the chosen gauge (\ref{unhatgauge}), this equation becomes:
\begin{eqnarray}
\partial_{\mu\nu}^2\frac{1+c_{,1}\square}{1+\frac{n}{2}c_{,1}\square}h^\alpha_\alpha - \square h_{\mu\nu} - \partial_{\mu\nu}^2h^\alpha_\alpha &&\nonumber\\
 + \left[\eta_{\mu\nu} +\frac{(n-1)(n-2)c_{,1}^2\square}{\left(1 + (n-1)c_{,1}\square\right)^2}\left(\partial_{\mu}\partial_{\nu}-\eta_{\mu\nu}\square\right) \right]\frac{1+(n-1)c_{,1}\square}{2+nc_{,1}\square}\square h^\alpha_\alpha&=&0\,,
\end{eqnarray}
again we study the trace equation and get:
\be
(n-2)\frac{1+2(n-1)c_{,1}\square}{\left(2+nc_{,1}\square\right)\cdot\left(1+(n-1)c_{,1}\square\right)}\cdot\square h^{\mu}_{\mu}=0\,.
\ee
The difference with the equation (\ref{eomnonhat}) is in the nonlocal operator $\frac{1}{2\left(1+(n-1)c_{,1}\square\right)}$ which is the price for making a change of variables ${\hat h}\to h$ with derivatives. This is always the case. If one makes a change of variables of the form $\phi\to Q\phi$ with some operator $Q$  in an action, then equations of motion $\frac{\delta S}{\delta (Q\phi)}=Q^{-1}\frac{\delta S}{\delta\phi}$ get multiplied by $Q^{-1}$.

Note that varying directly the action (\ref{sord}) instead of (\ref{nonlocal}) we would get yet another power of the non-local factor:
\be
(n-2)\frac{1+2(n-1)c_{,1}\square}{\left(2+nc_{,1}\square\right)\cdot\left(1+(n-1)c_{,1}\square\right)^2}\cdot\square h^{\mu}_{\mu}=0\,.
\ee
due to the overall factor of $\frac{1}{\left(1+(n-1)c_{,1}\square\right)}$  in the action. Once more, we see that it is very important to consistently treat the classes of variations and surface terms when dealing with such models.

Now the case with $c_{,1}\propto\frac{1}{\square}$ might look even more puzzling. Indeed, the operator $\frac{1}{1+(n-1)c_{,1}\square}$ in this case is just a number. However, we have seen that the spectrum in different pictures is different. The resolution is simple. When the function $C$ is a c-number, then the quadratic actions modulo the surface terms are numerically equal to each other in both pictures. It is just the domain of variations that has changed. However, when $c_{,1}\propto\frac{1}{\square}$ the second order correction to the relation between $h$ and $\hat h$ should be taken into account in the linear surface term in the initial action, and a part of it ceases to be a surface term after the change of variables to $h$. Indeed, the last term in (\ref{secondcur}) for $X_2$ is no longer a total derivative since the operator $\square c_{,2}$ is now a c-number. Therefore, even numerically the effective (after dropping the surface terms) Lagrangian densities are no longer coincident  in the two pictures.

\section{Discussion} \label{toy}

To illustrate some subtleties with boundary conditions and nonlocal field redefinitions in our considerations of equivalences between different theories, we 
consider a toy model with two scalar fields $\phi(x)$ and $\psi(x)$
\be
S=\int {\rm d}^n x\cdot \left(1+\psi(x)\right) \left(\square\phi(x) - (\partial\phi(x))^2\right)\,,
\ee
constrained by relation
\be
\phi=\psi+c_{,1}\left(\square\phi - (\partial\phi)^2\right)+c_{,1}\psi\square\phi+c_{,2}(\square\phi)^2+\ldots\,, 
\ee
which generalises the C-model relation for the two metrics being $1+\phi$ and $1+\psi$.

For the quadratic action in the $\phi$-picture, it is enough to solve for $\psi$ up to the linear order. We substitute
\be
\psi=\phi-c_{,1}\square\phi+\ldots\,,
\ee
and get
\be
S=-\int {\rm d}^n x\cdot \left(2 (\partial\phi)^2+c_{,1}(\square\phi)^2\right)\,,
\ee
which yields the equation of motion 
\be
2\square\phi-c_{,1}\square^2\phi=0\,,
\ee 
with higher order derivatives stemming from the derivative relation between the fields.

If we are not allowed to throw away the surface tems,  then the opposite transition requires solving for $\phi$ to second order:
\be
\phi=\frac{1}{1-c_{,1}\square}\left(\psi+c_{,1}\left(\psi\frac{\square}{1-c_{,1}\square}\psi+\left(\partial\frac{\psi}{1-c_{,1}\square}\right)^2\right)+c_{,2}\left(\frac{\square}{1-c_{,1}\square}\psi\right)^2+\ldots\right)\,.
\ee
The second order action is
\begin{multline}
S=\int d^4 x\left(\psi\frac{\square}{1-c_{,1}\square}\psi-\left(\partial\frac{\psi}{1-c_{,1}\square}\right)^2\right.\\
+\left.\frac{\square}{1-c_{,1}\square}\left(\psi+c_{,1}\left(\psi\frac{\square}{1-c_{,1}\square}\psi+\left(\partial\frac{\psi}{1-c_{,1}\square}\right)^2\right)+c_{,2}\left(\frac{\square}{1-c_{,1}\square}\psi\right)^2\right)\right)\,.
\end{multline}

Omitting the surface terms, we get the equation of motion
\be
\left(\frac{\square}{(1-c_{,1}\square)^2}+\frac{\square}{1-c_{,1}\square}\right)\psi=0\,,
\ee
which easily transforms to
\be
\frac{1}{1-c_{,1}\square}(2\square\phi-c_{,1}\square^2)\phi=0\,.
\ee
The difference between the two pictures amounts to the non-local operator which has been used for the change of variables from $\psi$ to $\phi$. This is precisely analogous to the relation we established for the C-theory in \ref{eom} between the $g$-picture in \ref{lnonhat} and $\hat{g}$-picture in \ref{lhat}.
Again, if $c_{,i}\propto\frac{1}{\square}$, the difference is much more profound, at least with the standard choice of boundary conditions. Indeed, in this case the term with $c_{,2}$ in the $\psi$-action is obviously not the surface one, and therefore the equations of motion differ by an extra non-trivial term.
\newline

The equivalence between different pictures is a tricky issue that requires exquisite care. We have nevertheless established that under reasonable assumptions, we can effectively regard wide classes of metric-affine gravities as nonlocal metric theories. We recovered infrared modifications alike previously studied models with cosmologically interesting phenomenology. However, in order to recover ultraviolet-complete ghost-free theories, nonlocalities needed to be implemented already in the fundamental relation between the affine and metric structures. 

%\section{Discussion}
%\label{discussion}

%There is a non-trivial problem of defining the correct boundary condition in the variational principle. Variations of which quantities must vanish at infinity? This is a particularly non-trivial problem in our case since the relation (\ref{relation}) contains derivatives. The equivalence between different pictures is a tricky issue that requires exquisite care.

%It is not obvious for the C and D models which one of the two metrics should be treated as a fundamental field. We have seen that the physical results generically depend on the chosen picture. Therefore an accurate definition of the variational principle is of vital importance for these models to have a well-defined physical content.

\acknowledgements{
AG is grateful to NORDITA for hospitality during his visits when this project has started and continued, and to the
Saint Petersburg State University travel grant 11.42.1388.2015; and also to the organisers of the ”Extended Theories
of Gravity” Programme at NORDITA in March 2015 and to the Dynasty Foundation for financial support in general
and in relation to participation in this programme in particular.}

\bibliography{refs}

\end{document}